\documentclass{Interspeech2024}

\interspeechcameraready

\title{SilentCipher: Deep Audio Watermarking}

\name[affiliation={1}]{Mayank Kumar}{Singh}
\name[affiliation={2}]{Naoya}{Takahashi}
\name[affiliation={1}]{Weihsiang}{Liao}
\name[affiliation={1}]{Yuki}{Mitsufuji}

\address{
  $^1$Sony AI, Japan\\
  $^2$Sony AI, Zurich}
\email{mayank.a.singh@sony.com, naoya.takahashi@sony.com,weihsiang.liao@sony.com,yuhki.mitsufuji@sony.com}


\usepackage{cite}
\usepackage{multirow}
\usepackage{makecell}
\usepackage{moresize}
\usepackage{changepage}

\begin{document}
\nocite{*}

\maketitle

\begin{abstract}
    In the realm of audio watermarking, it is challenging to simultaneously encode imperceptible messages while enhancing the message capacity and robustness. Although recent advancements in deep learning-based methods bolster the message capacity and robustness over traditional methods, the encoded messages introduce audible artefacts that restricts their usage in professional settings. In this study, we introduce three key innovations. Firstly, our work is the first deep learning-based model to integrate psychoacoustic model based thresholding to achieve imperceptible watermarks. Secondly, we introduce psuedo-differentiable compression layers, enhancing the robustness of our watermarking algorithm. Lastly, we introduce a method to eliminate the need for perceptual losses, enabling us to achieve SOTA in both robustness as well as imperceptible watermarking. Our contributions lead us to SilentCipher, a model enabling users to encode messages within audio signals sampled at 44.1kHz.
\end{abstract}

\section{Introduction}

Audio watermarking has seen widespread use in the past few decades, driven by the necessity to copyright online media and applications such as audio steganography. With advancements in the generation of synthetic audio samples that are indistinguishable from real samples, there's a growing demand for watermarking algorithms to identify audio synthesized using generative models like voice conversion\cite{starganv2_vc,our_IJCNN_SVC,our_EVC,our_robust_SVC, our_iterative_vc_asr}, text-to-speech\cite{mms_tts} and music generation\cite{musiclm}.

Traditional watermarking methods like LSB (Least Significant Bit) \cite{trad_LSB}, echo hiding\cite{trad_echo_hiding}, spread spectrum \cite{trad_spread_spec_1,trad_spread_spec_2,trad_spread_spec_3}, patchwork \cite{trad_patchwork}, and QIM (Quantization Index Modulation) \cite{trad_QIM} offer inaudible watermarks but are limited in encoding capacity and robustness against distortions. The watermarks generated by these algorithms can also be removed by systematic attacks as shown in \cite{watermark_attack}.


To address these limitations, several works have turned to deep learning models to enhance robustness and capacity \cite{hideandspeak, robustdnn, wavmark, our_watermarking}. However, unlike the traditional algorithms which make use of expert knowledge to ensure that the introduced artefacts are inaudible, current state-of-the-art deep learning models do not incorporate such constraints. Recently, \cite{meta_recent}, uses a perceptual loss which minimizes localized frequencies between the source and the watermarked audio. Although this incorporates rudimentary psychoacoustic knowledge, it does not necessarily ensure that the message encoded by their algorithm adheres to the frequency masking threshold, thus allowing their model to introduce audible artefacts.


In this study, we find that deep learning-based algorithms introduce audible artefacts in watermarked samples, which we attribute to the lack of integrating psychoacoustic knowledge, particularly noticeable in band-limited signals. This makes them unreliable to use in professional setups where the original content preservation is of paramount importance. Although \cite{wavmark} mentions the audible artefacts in silent regions of the watermarked audio and proposes an SDR based selective message encoding, during our evaluations we found that it is not just the silent regions but also band-limited regions that have perceptible artefacts which make the selective message encoding task non-trivial. We build upon \cite{hideandspeak,wavmark} and introduce a new model, SilentCipher, which is more robust to distortions like audio compression, time-jittering and additive white noise while encoding imperceptible messages even for the challenging case of band-limited audio signals.
While \cite{wavmark} finds it difficult to scale their model beyond a 16kHz sampling rate due to limitations in model size and computation, we address this issue by extending our model to accommodate a sampling rate of 44.1kHz, suitable for professional applications, with minimal computational and memory requirements.


Overall, the contributions of our paper are summarized as follows. We propose SilentCipher, the first deep learning based model which takes inspiration from psychoacoustic model based thresholding to realize inaudible message encoding while achieving higher message bit-rate and accuracy than state-of-the-art methods on various distortions. We also propose a method to incorporate non-differential audio compression algorithms while training deep learning models that allows us to robustify SilentCipher to popular compression methods. We further introduce a method that removes the need for perceptual losses while allowing us to have the flexibility of determining a user-controlled lower bound for SDR during inference. Further, we extend SilentCipher to audio signals at a sampling frequency of 44.1kHz, allowing its usage in more practical and professional applications. Demo audio samples can be found online at \footnote{\href{https://interspeech2024.github.io/silentcipher/}{https://interspeech2024.github.io/silentcipher/}}.

\section{Related Works}


Previous works on audio watermarking fall into two categories: traditional methods such as LSB (Least Significant Bit)\cite{trad_LSB}, echo hiding\cite{trad_echo_hiding}, spread spectrum\cite{trad_spread_spec_1,trad_spread_spec_2,trad_spread_spec_3}, patchwork\cite{trad_patchwork}, and QIM (Quantization Index Modulation)\cite{trad_QIM}, which have been prominent for decades, and deep learning-based methods, which have demonstrated superior encoding capacity and robustness compared to traditional approaches.

The proliferation of algorithms enhancing sample quality in voice conversion \cite{starganv2_vc,our_IJCNN_SVC,our_EVC,our_robust_SVC,our_iterative_vc_asr}, text-to-speech, and music generation has heightened interest in bolstering the encoding capacity and robustness of watermarking techniques. While a recent deep learning-based audio watermarking method \cite{wavmark} claims superior signal to distortion ratio (SDR) compared to traditional methods like \cite{swesterfeld}, promising improved imperceptibility, our extensive experiments reveal that a higher SDR doesn't necessarily guarantee imperceptible artifacts to human perception. This discrepancy arises from the neglect of psychoacoustics, which determines human perception of artefacts in presence of the carrier.



While deep learning-based methods often boast higher SDR \cite{wavmark}, they typically cannot establish a lower bound for SDR on a per-utterance basis, which is crucial for practical applications where tolerance for audible artifacts is minimal. In our model, we introduce a novel approach to parameterize and modify the lower bound for SDR without requiring additional training during inference.


\begin{figure}[t]
  \centering
  \includegraphics[width=1.02\linewidth]{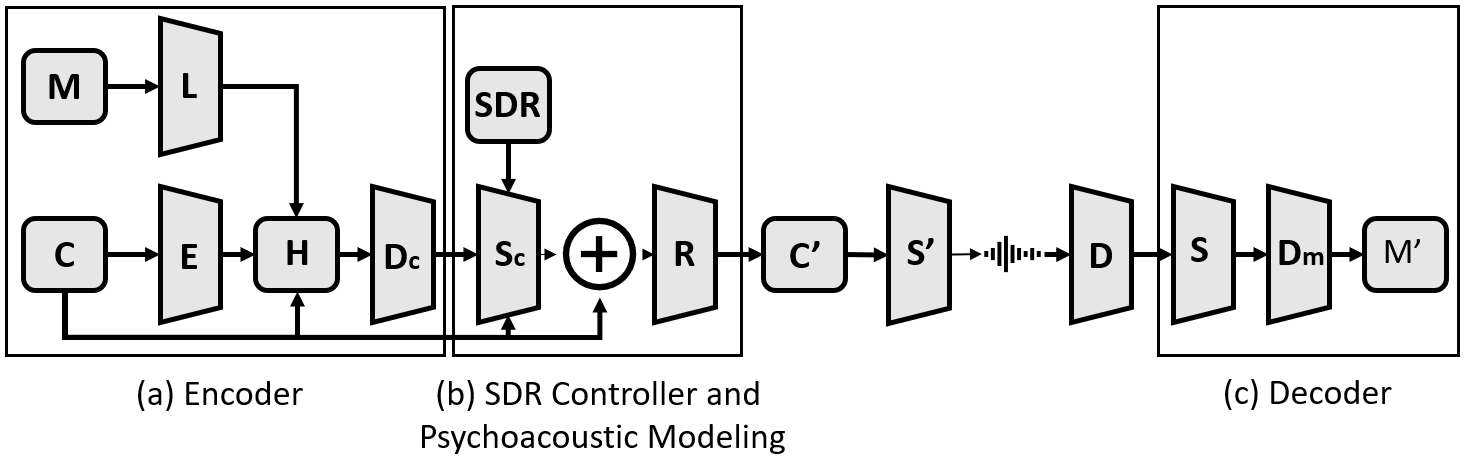}
  \caption{The model architecture consists of the message transformation network, $L$, encoder network, $E$, the carrier decoder network $D_{c}$, the scaler operation, $S_{c}$, the ReLU operation, $R$, and the message decoder $D_{m}$.}
  \label{fig:model}
  \vspace{-5mm}
\end{figure}

\section{Proposed Method}

Our model architecture, which is based upon the works of \cite{hideandspeak}, is shown as in figure \ref{fig:model}. The message tokens, $M\in\mathbb{R}^{1\times 1\times T}$, are projected to learnable embeddings using $L$ to get $M_{e}(M)\in\mathbb{R}^{1\times N\times T}$, where N is the fft size. The encoder gets the carrier, $C\in\mathbb{R}^{1\times N\times T}$ which is the magnitude of the STFT of the source audio signal, and outputs $E(C)$ which is concatenated along the channel axis with $C$ and $M_{e}(M)$ to get $H$. The carrier decoder, represented as $D_{c}$, takes input $H$ and produces an output denoted as $D_c(H)$. We apply a mask to $D_c(H)$ such that the message is confined to the lower half frequencies of the spectrogram.

\subsection{Lower bound of SDR}

As illustrated in Figure \ref{fig:model} (b), instead of using a perceptual loss to minimize the distortions between the embedded carrier and the source carrier as proposed in \cite{hideandspeak,our_watermarking}, we introduce a novel method that guarantees a lower bound of the SDR by scaling the additive message signal, $D_{c}(H)\in\mathbb{R}^{1\times N\times T}$, by a user-provided factor, $\alpha$, of the source carrier as defined by the following equation,

\setlength{\abovedisplayskip}{-5pt}
\setlength{\belowdisplayskip}{4pt}

\begin{equation}
  \label{equation:eq1}
ReLU(C + W)
\end{equation}

We get the embedded carrier, $C'$, by adding $S_{c}(D_{c}, C, \alpha)$ to $C$. As shown in Eq. \ref{equation:eq2} and Eq. \ref{equation:eq3}, by scaling the message signal using Eq. \ref{equation:eq1}, the perceptual loss, $L_{p}$, given by $||C' - C||_{2}$ as well as the SDR are independent of the network parameters.

\begin{equation}
    \label{equation:eq2}
    \begin{split}
        L_{p} &= ||C' - C||_{2} = ||S_{c}(D_{c}(H), C, \alpha)||_{2}\\
        &= ||C||_2\times 10^{-\alpha/20}
    \end{split}
\end{equation}

The application of the ReLU activation function, $R$, and the inverse STFT function, denoted as $S'$, to the signal $C'$ results in the watermarked carrier, $\Tilde{c}$. For $S'$, the phase is taken without modification from the original signal. $R$, thresholds the message spectrogram to an upper limit of the carrier spectrogram which further enhances the SDR, enabling us to establish a user-controlled lower threshold for SDR post-embedding.

\begin{equation}
    \label{equation:eq3}
    \begin{split}
        \text{SDR} &= 20 \log (||C||_2 / ||C' - C||_2) \\
        &= 20 \log (||C||_2 / (||C||_{2}\times 10^{-\alpha/20}) \\
        &= 20 \log (10^{\alpha/20}) = \alpha
    \end{split}
\end{equation}

\subsection{Making the artefacts imperceptible}

Inspired by psychoacoustics, we aim to ensure the imperceptibility of the added watermark to human listeners. Psychoacoustic research indicates that humans struggle to discern a low-energy audio signal amidst a high-energy one when their frequencies are closely aligned \cite{freq_mask}. Motivated by this insight, we enforce two conditions: firstly, we set the phase of the embedded watermark to $\pi$ relative to the original carrier for each frequency bin; secondly, we constrain the magnitude of each frequency bin to be less than or equal to that of the original signal. This prompts the network to allocate more energy to frequency bins with higher magnitudes in the original signal, while assigning less energy to those with lower magnitudes. Empirical experimentation validates the imperceptibility of the audio watermark to human auditory perception.


We enforce the constraint of the phase of the embedded watermark being $\pi$ for each frequency bin by ensuring that $S_{c}$ is non-positive. To this end, during the scaling operation, we take the absolute of $D_c(H)$ and multiply the equation by -1 as depicted in Equation \ref{equation:eq1}. This, combined with the application of the ReLU function, $R$, guarantees that the magnitude spectrogram of the embedded watermark for every frequency bin is smaller than that of the original signal, while simultaneously preventing negative spectral magnitudes.

\begin{table*}[!th]
  \centering
  \belowrulesep = 0mm
  \ssmall
  \caption{Objective Test Results. We compare the baselines using objective test scores by simulating various attacks. Runtime: the reciprocal of the average time taken by each model for encoding 1 second of audio. GN: additive Gaussian noise of 40dB, 50C: randomly cropping the audio by 50\% of its total duration, EQ: random equalization, MX: mixing the encoded waveform with speech at -15dB, Q: 16-bit floating-point Quantization, TJ: time-jittering and RS: random resampling from 6.4kHz to 16kHz.}
  \label{tab:obj_results}
  \vspace{-1mm}
 \begin{adjustwidth}{-0.1cm}{-1cm}
  \begin{tabular}{ c@{\hspace{2mm}} c@{\hspace{2mm}} c@{\hspace{2mm}} c@{\hspace{2mm}} c@{\hspace{2mm}} c@{\hspace{2mm}} c@{\hspace{2mm}} c@{\hspace{2mm}} c@{\hspace{2mm}} c@{\hspace{2mm}} c@{\hspace{2mm}} c@{\hspace{2mm}} c@{\hspace{2mm}} c@{\hspace{2mm}} c@{\hspace{2mm}} c@{\hspace{2mm}} c@{\hspace{2mm}} c@{\hspace{2mm}} c@{\hspace{2mm}} c@{\hspace{2mm}} c@{\hspace{2mm}} c@{\hspace{2mm}} c@{\hspace{2mm}} c@{\hspace{2mm}}}
    \toprule
    \\[-2mm]
    \textbf{\makecell{Dur.\\(secs)}} & \textbf{\makecell{Models}} & \textbf{\makecell{Run\\time}} & \textbf{\makecell{Mean\\Acc.}} & \textbf{\makecell{No\\attack}} & \textbf{\makecell{GN}} & \textbf{\makecell{50C}} & \textbf{\makecell{EQ}} & \textbf{\makecell{MX}} & \textbf{\makecell{Q}} & \textbf{\makecell{TJ}} & \textbf{\makecell{RS}} & \textbf{\makecell{MP3\\64}} & \textbf{\makecell{MP3\\128}} & \textbf{\makecell{MP3\\256}} & \textbf{\makecell{OGG\\64}} & \textbf{\makecell{OGG\\128}} & \textbf{\makecell{OGG\\256}} & \textbf{\makecell{AAC\\64}} & \textbf{\makecell{AAC\\128}} & \textbf{\makecell{AAC\\256}} \\
    \midrule
    \rule{0pt}{2mm}
    \multirow{4}{3mm}{6}      & wavmark & $89$ & $96.16$ & $98.64$ & $98.61$ & $98.18$ & $\textbf{100}$  & $\textbf{100}$ & $\textbf{100}$ & $97.71$ & $97.5$ & $95.3$ & $95.52$ & $95.52$ & $95.52$ & $95.52$ & $95.52$ & $\textbf{95.52}$ & $95.52$ & $95.52$\\
     & Audiowmark & $146$ & $25.89$       & $32.33$ & $24.77$ & $1.22$ & $25.11$ & $26.11$ & $31.77$ & $29.88$ & $32.55$ & $17.55$ & $30.44$ & $32.44$ & $32.33$ & $32.33$ & $32.33$ & $8.11$ & $25.55$ & $31.77$ \\
     & robustdnn & $205$ & $69.97$         & $\textbf{100}$ & $\textbf{100}$ & $65.22$ & $\textbf{100}$ & $93.11$ & $\textbf{100}$ & $\textbf{100}$ & $20.88$ & $93.11$ & $93.11$ & $93.11$ & $20$ & $93.11$ & $93.11$ & $16.22$ & $17.33$ & $17.33$\\
     & SC-16 & $\textbf{1302}$ & $\textbf{98.93}$  & $\textbf{100}$ & $\textbf{100}$ & $\textbf{99.88}$ & $92.55$ & $\textbf{100}$ & $\textbf{100}$ & $\textbf{100}$ & $\textbf{100}$ & $\textbf{96}$ & $\textbf{100}$ & $\textbf{100}$ & $\textbf{100}$ & $\textbf{100}$ & $\textbf{100}$ & $94.55$ & $\textbf{100}$ & $\textbf{100}$ \\
    \bottomrule
    \rule{0pt}{2mm}
    \multirow{4}{3mm}{12}     & wavmark & $-$ & $97.68$          & $\textbf{100}$ & $97.01$ & $99.98$ & $97.51$ & $97.51$ & $97.51$ & $99.31$  & $99.97$ & $96.12$ & $96.12$ & $96.12$ & $96.12$ & $96.12$ & $96.12$ & $96.12$ & $96.12$ & $96.12$\\
    & Audiowmark & $-$ & $78.98$       & $89.09$ & $78.62$ & $27.41$ & $79.29$ & $80.40$ & $79.44$ & $79.44$ & $79.44$ & $79.44$ & $79.44$ & $79.44$ & $79.44$ & $79.44$ & $79.44$ & $76.11$ & $79.44$ & $79.44$ \\
    & robustdnn & $-$ & $70.01$         & $\textbf{100}$ & $\textbf{100}$ & $68.66$ & $\textbf{100}$ & $93.55$ & $\textbf{100}$ & $\textbf{100}$ & $21.28$ & $93.55$ & $93.55$ & $93.55$ & $19.44$ & $93.55$ & $93.55$ & $17.33$ & $17.33$ & $18$ \\
    & SC-16 & $-$ & $\textbf{99.59}$  & $\textbf{100}$ & $\textbf{100}$ & $\textbf{100}$ & $96.22$ & $\textbf{100}$ & $\textbf{100}$ & $\textbf{100}$ & $\textbf{100}$ & $\textbf{98.22}$ & $\textbf{100}$ & $\textbf{100}$ & $\textbf{100}$ & $\textbf{100}$ & $\textbf{100}$ & $\textbf{99.11}$ & $\textbf{100}$ & $\textbf{100}$ \\
    \bottomrule
    \rule{0pt}{2mm}
    \multirow{4}{3mm}{24}     & wavmark & $-$ & $\textbf{100}$          & $\textbf{100}$ & $\textbf{100}$ & $\textbf{100}$ & $\textbf{100}$ & $\textbf{100}$ & $\textbf{100}$ & $\textbf{100}$ & $\textbf{100}$ & $\textbf{100}$ & $\textbf{100}$ & $\textbf{100}$ & $\textbf{100}$ & $\textbf{100}$ & $\textbf{100}$ & $\textbf{100}$ & $\textbf{100}$ & $\textbf{100}$ \\
    & Audiowmark & $-$ & $78.92$       & $89.44$ & $79.44$ & $74.44$ & $79.44$ & $79.44$ & $82.18$ & $80.18$ & $82.18$ & $79.30$ & $87.97$ & $89.31$ & $89.09$ & $89.09$ & $88.86$ & $53.25$ & $87.53$ & $89.09$ \\
    & robustdnn & $-$ & $73.02$         & $\textbf{100}$ & $\textbf{100}$ & $64.28$ & $\textbf{100}$ & $\textbf{100}$ & $\textbf{100}$ & $\textbf{100}$ & $22$ & $\textbf{100}$ & $\textbf{100}$ & $\textbf{100}$ & $20.73$ & $\textbf{100}$ & $\textbf{100}$ & $20.71$ & $20.71$ & $20.71$ \\
    & SC-16 & $-$ & $99.82$  & $\textbf{100}$ & $\textbf{100}$ & $\textbf{100}$ & $98.88$ & $\textbf{100}$ & $\textbf{100}$ & $\textbf{100}$ & $\textbf{100}$ & $98.88$ & $\textbf{100}$ & $\textbf{100}$ & $\textbf{100}$ & $\textbf{100}$ & $\textbf{100}$ & $99.44$ & $\textbf{100}$ & $\textbf{100}$ \\
    \bottomrule
    \rule{0pt}{3mm}
    24     & SC-44 & $\textbf{312}$ & $\textbf{99.96}$          & $\textbf{100}$ & $\textbf{100}$ & $\textbf{100}$ & $\textbf{100}$ & $\textbf{100}$ & $\textbf{100}$ & $\textbf{100}$  & $\textbf{100}$ & $\textbf{100}$ & $\textbf{100}$ & $\textbf{100}$ & $\textbf{100}$ & $\textbf{100}$ & $\textbf{100}$ & $99.44$ & $\textbf{100}$ & $\textbf{100}$\\
    \bottomrule
  \end{tabular}
  \normalsize
  \vspace{-3mm}
\end{adjustwidth}
\end{table*}
\subsection{Making SilentCipher robust to distortions}

To enhance SilentCipher's resilience against attacks, we incorporate various differentiable distortions during training, including additive Gaussian noise, time-jittering, and random equalization of frequency bands. However, while these distortions broaden the model's robustness to a diverse range of attacks, the constraint of only employing differentiable distortions limits its effectiveness against non-differentiable audio compression algorithms. Although training a differentiable substitute for the MP3 compression algorithm using deep neural networks is conceivable, practical implementation reveals insufficient precision in reproducing the intricacies of the MP3 compression algorithm.
To address these limitations and bolster our model's robustness against compression algorithms like MP3, OGG, and AAC, we introduce a pseudo-differentiable compression layer. During the forward pass of gradient descent, this layer applies compression, while during the backward pass, gradients bypass this layer. Empirical evidence presented in Section \ref{section:Results} demonstrates that the introduction of the pseudo-differentiable compression layer effectively fortifies our proposed model against various audio compression algorithms.
\\
Additionally, to make SilentCipher robust against resampling attacks up to half of the original sampling rate, we mask $D_{c}(H)$ to ensure that it is non-zero only for frequency bins below half of the maximum frequency. The message decoder, $D_{m}$, gets the magnitude of the STFT of the embedded signal after applying the distortions and outputs the message predictions, $M'$.


In \cite{hideandspeak,our_watermarking}, the two losses, message classification loss and perceptual loss, are optimized together and a crude weighing hyper-parameter is used to control the trade-off between accuracy and SDR. Because of the competing nature of this framework, it introduces difficulties for the convergence of the model. Unlike \cite{hideandspeak,our_watermarking}, SilentCipher is trained by only optimizing the message the cross-entropy loss between $M$ and $M'$. We can omit the perceptual loss because of its independence to the parameters of the network as shown in Eq. \ref{equation:eq2} and \ref{equation:eq3}.


By repeating the message along the time-axis during carrier encoding, we reduce the error through averaging the prediction across the repeated segments. To make our model robust to time-cropping, we adopt a brute-force method to determine the phase which maximizes the frequency of the mode of the set of predicted message characters separated by an interval equal to the message length. This allows us to remove the adverse-effects of time-cropping. To find the end-position of each message segment, we allocate an extra bit and message character for the end-token. It is trivial to detect whether the audio is watermarked by checking if the maximum repeating frequency of any message character is less than a certain threshold.
\section{Experiments}
\label{section:Experiments}

Two versions of our model, SilentCipher-16k (SC-16) and SilentCipher-44k (SC-44), are trained on audio with sampling frequencies of 16kHz and 44.1kHz, respectively. As the public implementations of WavMark and RobustDNN are limited to watermarking audio at a sampling frequency of 16kHz, to ensure fairness in evaluation, we compare SC-16 with the baseline models across various audio domains including speech, music, vocals, soap operas, and instrumentals. Evaluations are conducted on durations of 6, 12, and 24 seconds to assess the models' ability to leverage longer utterances for improved robustness against distortions. During training, SC-16 and SC-44 uniformly sample binary messages of lengths 32 bits and 40 bits, respectively, which are repeated to match the number of spectrogram frames. For evaluation purposes, repeating binary messages of lengths recommended in the baseline models are embedded.
\\
\textbf{Datasets}\hspace{2mm}  For the speech dataset, we use the subset of non-singing voice from VCTK\cite{vctk}, NUS-48E\cite{nus} and NHSS\cite{nhss} comprising of 44 hours, 1 hour and 2.25 hours of audio, respectively.
For the singing vocals datasets, we use the subset of singing voice from NUS-48E, NHSS and internal vocals dataset comprising of 2, 4.75 and 90 hours of audio, respectively.
For the music and instrumental dataset, we use MUSDB18\cite{musdb} with an audio duration of 6.5 hours. 
For soap operas, we use an internal dataset with a total duration of 321 hours which consists of game shows, soap operas and kids cartoon. The train, validation and testing set are split in the ratio 0.8:0.1:0.1. We process the data for SilentCipher-16k by extracting the STFT of the waveform, with the size of the Fourier transform being 2048, window length being 2048 and hop length being 1024. For SilentCipher-44k, the STFT parameters are different with the size of the Fourier transform being 4096, window length being 4096 and hop length being 2048. A Hann window was used to calculate the STFT.
\begin{table}[t]
  \centering
  \belowrulesep = 0.75mm
  \footnotesize
  \caption{Subjective Test Results comparing our model against the baselines. A higher score for mean inaudibility corresponds with inaudible artefacts in the embedded carrier. We also report the 95\% Confidence Intervals (CI).}
  \label{tab:subjective_test}
  \begin{tabular}{ c c c c }
    \toprule
    Model & Mean Inaudibility & CI low & CI high \\
    \toprule
    SilentCipher & $1$ & $0.87$ & $1$\\
    AudioWMark & $1$ & $0.87$ & $1$\\
    Wavmark & $0$ & $0$ & $0.12$\\
    RobustDNN & $0$ & $0$ & $0.12$\\
    \bottomrule
  \end{tabular}
  \normalsize
  
\vspace{-5mm}
\end{table}
\\
\textbf{Training}\hspace{2mm} All of our models were trained using the Adam optimizer with a learning rate of $10^{-3}$ for a total of 80000 iterations. The audio duration during training is fixed to be 12 seconds. We implement the encoder and decoder as per the hyper-parameters provided in \cite{hideandspeak}. One of the distortions is selected uniformly from a set that encompasses additive Gaussian noise, time-jittering, random equalization of frequency bands, and audio compression algorithms. If the chosen distortion involves compression algorithms, we uniformly sample the bit-rate from 64kbps, 128kbps, and 256kbps.We also uniformly sample the type of compression algorithm from MP3, OGG, and AAC.

\begin{table*}[!th]
  \centering
  \belowrulesep = 0mm
  \ssmall
  \caption{Ablation Test Results. SubEval refers to the Subjective evaluation results which is represented in the format [mean;lower limit of 95\% confidence interval; upper limit of 95\% confidence interval] where a higher value corresponds to inaudible artefacts.}
  \label{tab:ablation}
  \begin{adjustwidth}{-0.1cm}{-1cm}
  \begin{tabular}{ l@{\hspace{2mm}} c@{\hspace{2mm}} c@{\hspace{2mm}} c@{\hspace{2mm}} c@{\hspace{2mm}} c@{\hspace{2mm}} c@{\hspace{2mm}} c@{\hspace{2mm}} c@{\hspace{2mm}} c@{\hspace{2mm}} c@{\hspace{2mm}} c@{\hspace{2mm}} c@{\hspace{2mm}} c@{\hspace{2mm}} c@{\hspace{2mm}} c@{\hspace{2mm}} c@{\hspace{2mm}} c@{\hspace{2mm}} c@{\hspace{2mm}} c@{\hspace{2mm}} c@{\hspace{2mm}}}
    \toprule
    \\[-2mm]
    \textbf{\makecell{Models}} &  \textbf{\makecell{SDR $\uparrow$}} & \textbf{\makecell{SubEval $\uparrow$}} & \textbf{\makecell{Mean\\Acc.}} & \textbf{\makecell{No\\attack}} & \textbf{\makecell{GN}} & \textbf{\makecell{50C}} & \textbf{\makecell{EQ}} & \textbf{\makecell{MX}} & \textbf{\makecell{Q}} & \textbf{\makecell{TJ}} & \textbf{\makecell{RS}} & \textbf{\makecell{MP3\\64}} & \textbf{\makecell{MP3\\128}} & \textbf{\makecell{MP3\\256}} & \textbf{\makecell{OGG\\64}} & \textbf{\makecell{OGG\\128}} & \textbf{\makecell{OGG\\256}} & \textbf{\makecell{AAC\\64}} & \textbf{\makecell{AAC\\128}} & \textbf{\makecell{AAC\\256}} \\
    \midrule
    \rule{0pt}{2mm}
    SC-16   & $47.24$ & $1;0.87;1$ & $99.03$ & $100$ & $100$ & $100$ & $96.22$ & $100$ & $100$ & $100$ & $\textbf{100}$ & $100$ & $100$ & $100$ & $100$ & $100$ & $100$ & $99.11$ & $100$ & $100$ \\
    - half-band  & $47.42$ & $\textbf{1;0.87;1}$ & $99.11$ & $100$ & $100$ & $100$ & $97.33$ & $100$ & $100$ & $100$ & $98.66$ & $100$ & $100$ & $100$ & $100$ & $100$ & $100$ & $100$ & $100$ & $100$\\
    - neg msg  & $47$ & $0;0;0.13$ & $\textbf{99.33}$ & $100$ & $100$ & $100$ & $99.88$ & $100$ & $100$ & $100$ & $0$ & $\textbf{100}$ & $\textbf{100}$ & $\textbf{100}$ & $100$ & $100$ & $100$ & $100$ & $100$ & $100$ \\
    - compression  & $\textbf{47}$ & $0.19;0.06;0.38$ & $76.15$ & $100$ & $100$ & $100$ & $100$ & $100$ & $100$ & $100$ & $0$ & $0$ & $0$ & $0$ & $100$ & $100$ & $100$ & $100$ & $100$ & $100$ \\
    - controlled SDR & $34.23$ & $0;0;0.13$ & $75.83$ & $100$ & $100$ & $100$ & $95.11$  & $100$ & $100$ & $100$ & $0$ & $0$ & $0$ & $0$ & $100$ & $100$ & $100$ & $99.33$ & $99.33$ & $99.33$ \\
    \bottomrule
  \end{tabular}
  \end{adjustwidth}
  \normalsize
\vspace{-3mm}
\end{table*}

\textbf{Baselines}\hspace{2mm} We conduct comparisons with WavMark \cite{wavmark}, RobustDNN \cite{robustdnn}, and AudioWmark \cite{swesterfeld}. Given that all baseline models support a 16kHz sampling rate, we employ our SC-16 model for the comparision. The bits-per-second (BPS) embedded by WavMark, AudioWmark, RobustDNN and SC-16 are 32, 20, 1.3 and 32, respectively, while the SDR is 38.27dB, 33.29dB, 18.49dB, 47.24dB, respectively.
\\
\textbf{Analysis of individual components}\hspace{2mm} Beginning with SC-16, we remove the constraint limiting the message to frequency bands below 4kHz (half-band). Subsequently, we eliminate the requirement for the message to be negative and conform to the frequency masking threshold based on psychoacoustics(neg. msg). Following this, we remove the pseudo-differentiable compression layers from $D$ (compression), and subsequently, the SDR controller block (controlled SDR) shown in Fig \ref{fig:model} (b).

\section{Results}
\label{section:Results}

\textbf{Objective Results}\hspace{2mm}To objectively evaluate SC-16 with the baselines in various scenarios, we distort the encoded signal and note the accuracy of the models for decoding the message for varying utterance duration as depicted in table \ref{tab:obj_results}. The distortions that are applied are additive Gaussian noise of 40dB (GN), random crop of the signal by 50\%(50C), applying random band-limited equalization of 15dB at 35Hz, 200Hz, 1000Hz and 4000Hz (EQ), mixing the watermarked carrier with a speech signal (MX) which is sampled uniformly from the NUS and NHSS datasets, 16-bit floating-point quantization(Q), time-jittering (TJ), random resampling of the audio by uniformly sampling from a sampling frequency of 6.4kHz to 16kHz (RS), MP3, OGG and AAC compression with bit-rates of 64, 128 and 256kbps. As can be seen from the Table \ref{tab:obj_results}, SC-16 overall outperforms the baselines while having an higher average SDR. SC-16 also runs 1302 times faster than realtime which is more than 6 and 9 times faster than RobustDNN and wavmark, respectively. We also note that we achieve superior results while using just 372 hours of training data compared to 5k hours used by WavMark. SC-44 watermarks audio at a sampling rate of 44.1kHz while still being faster than the baselines. For runtime calculations, AudioWMark is evaluated using an  Intel(R) Xeon(R) CPU E5-2620 v4 @ 2.10GHz, as the public implementation of AudioWMark does not support GPU implementation, whereas the other models are evaluated using 1 NVIDIA A100 GPU.
\\
\textbf{Subjective Results}\hspace{2mm}We conducted a subjective evaluation involving expert audio engineers to assess the detectability of artifacts in the encoded signal. Each evaluator was presented with two audio samples: one original and one encoded. They were asked to indicate whether they perceived an audible difference between the two samples in a binary format. On average, each evaluator assessed a total of 25 samples, with two samples serving as positive anchors and two as negative anchors. The positive anchors, selected from our ablation study utilizing perceptual loss, were chosen to ensure a discernible difference between the pairs. The negative anchors lacked any encoded watermark and consisted of identical audio samples. Evaluations rendered by engineers unable to provide accurate assessments of the anchors were omitted from the analysis, resulting in a subset of 12 competent evaluators.

For audio samples with energy spread across the frequency spectrum, all watermarking algorithms successfully encode completely inaudible messages. However, samples with band-limited energy are susceptible to audible watermarks added by the models. To address this, we filter out samples with contiguous segments longer than 1 second, where over 90\% of their energy is concentrated in less than 50\% of the lower frequency bands. The frequency of this condition being satisfied for our evaluation dataset is 19\%. The subjective evaluation results are reported in Table \ref{tab:subjective_test}. We find that the artefacts introduced by wavmark and robustDNN are always audible to the evaluators while they are unable to distinguish any artefacts in the watermarked samples from SilentCipher and AudioWMark.

From the subjective evaluation, we also found that wavmark occasionally alters the original content of the signal whereas our model always preserves the original content because of the strong restrictions imposed on it.
\\\textbf{Analysis of individual components}\hspace{2mm}We are able to validate the effectiveness of having controllable message energy, pseudo-differentiable compression layers as well as the negative message constraint from Table \ref{tab:ablation}.
We observe that having a user-controlled lower bound for the SDR improves the convergence of the model and achieves a higher SDR of 47dB while improving the subjective evaluation results as well. Introducing pseudo differentiable compression algorithm makes the model robust to mp3 algorithms. Restricting the message to be negative makes the artefacts inaudible as shown by the subjective tests, while introducing half-band makes the model further robust to random resampling attacks. 



\begin{figure}[t]
  \centering
  \includegraphics[width=\linewidth]{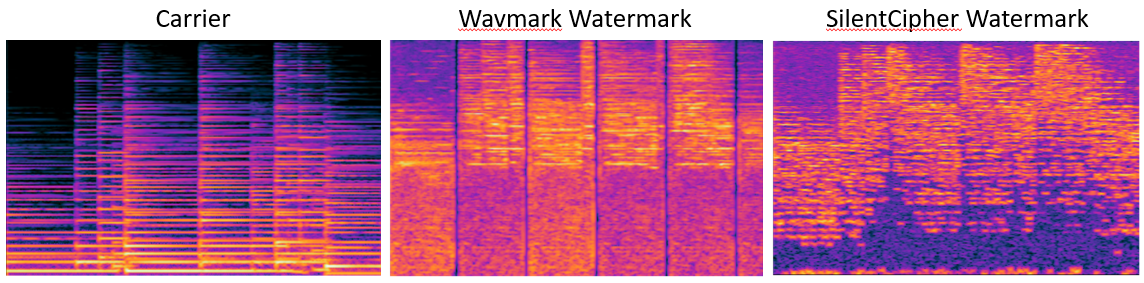}
  \caption{Visual analysis of the watermarks}
  \label{fig:visualize}
  \vspace{-6mm}
\end{figure}

\section{Observations}

From a visual examination of the normalized message magnitude spectrogram for wavmark and SilentCipher, as shown in Figure \ref{fig:visualize}, we also find that the message encoded by SilentCipher distributes its energy into frequency bands similar to that of the carrier, whereas the message encoded by wavmark has periodic characteristics and doesn't strongly correspond with the underlying carrier. This makes it difficult to design attacks which can remove the watermark encoded by SilentCipher compared to that of wavmark.

\section{Conclusions}

We introduce SilentCipher, a novel watermarking algorithm that, through extensive objective and subjective assessments, we validate as the SOTA in terms of inaudible message encoding and robustness against distortion. This is realized through the integration of techniques like pseudo-differentiable layers, user-controlled lower bounds for SDR, and ensuring that the message adheres to psychoacoustic frequency masking thresholds. Future endeavors entail enhancing our model's resilience to stronger mixing and over-the-air transmission distortions.

\pagebreak
\bibliographystyle{IEEEtran}
\bibliography{mybib}

\end{document}